\documentstyle[aps,epsfig,twocolumn]{revtex}
\begin{document}
\draft
\title{
A Measurement of the Coulomb Dissociation of $^8$B \\
at 254 MeV/nucleon and  the
$^8$B Solar Neutrino Flux
}
\author{
N.~{\sc Iwasa}$^{1,2}$,
F.~{\sc Bou\'e}$^{1,3}$,
G.~{\sc Sur\'owka}$^{1,4}$,
K.~{\sc S\"ummerer}$^{1}$,
T.~{\sc Baumann}$^{1}$,
B.~{\sc Blank}$^{3}$,
S.~{\sc Czajkowski}$^{3}$,
A.~{\sc F\"orster}$^{5}$,
M.~{\sc Gai}$^{6}$,
H.~{\sc Geissel}$^{1}$,
E.~{\sc Grosse}$^{7}$,
M.~{\sc Hellstr\"om}$^{1}$,
P.~{\sc Koczon}$^{1}$,
B.~{\sc Kohlmeyer}$^{8}$,
R.~{\sc Kulessa}$^{4}$,
F.~{\sc Laue}$^{5}$,
C.~{\sc Marchand}$^{4}$,
T.~{\sc Motobayashi}$^{9}$,
H.~{\sc Oeschler}$^{5}$,
A.~{\sc Ozawa}$^{2}$,
M.~{\sc S.~Pravikoff}$^{3}$,
E.~{\sc Schwab}$^{1}$,
W.~{\sc Schwab}$^{1}$,
P.~{\sc Senger}$^{1}$,
J.~{\sc Speer}$^8$,
C.~{\sc Sturm}$^5$,
A.~{\sc Surowiec}$^{1}$,
T.~{\sc Teranishi}$^{2}$,
F.~{\sc Uhlig}$^{5}$,
A.~{\sc Wagner}$^{5}$,
W.~{\sc Walus}$^{4}$, 
and 
C.A.~{\sc Bertulani}$^{10}$
}
\address{
$^1$ Gesellschaft f\"ur Schwerionenforschung m.b.H. (GSI),
D-64291 Darmstadt, Germany.\\
$^2$ RIKEN (Institute of Physical and Chemical Research),
Wako, Saitama 351-0198, Japan.\\
$^3$ Centre d'Etudes Nucl\'eaires de Bordeaux-Gradignan,
F-33175 Gradignan, France.\\
$^4$ Jagellonian University, PL-30-059 Krakow, Poland.\\
$^5$ Technical University of Darmstadt,
D-64289 Darmstadt, Germany.\\
$^6$ University of Connecticut,
Storrs, CT 06269-3046, U.S.A.\\
$^7$ Forschungszentrum Rossendorf,
D-01314 Dresden, Germany.\\
$^8$ Marburg University,
D-35032 Marburg, Germany.\\
$^9$ Rikkyo University, Toshima, Tokyo 171, Japan.\\
$^{10}$ Instituto de Fisica, Universidade Federal do Rio de Janeiro,  
21945-970 RJ, Brazil.\\
}
\maketitle
\newpage
\begin{abstract}
   We have measured the Coulomb dissociation of $^8$B into $^7$Be and proton
at 254~MeV/nucleon using a large-acceptance focusing spectrometer.
The astrophysical $S_{17}$ factor for the 
$^7$Be(p,$\gamma$)$^8$B reaction at $E_{\rm c.m.}=0.25-2.78$~MeV is deduced
yielding $S_{17}(0)=20.6 \pm 1.2$~(exp.) $\pm~1.0$~(theo.)~eV-b.
   This result agrees with the presently adopted zero-energy
$S_{17}$ factor  obtained in direct-reaction measurements and with
the results of other Coulomb-dissociation studies performed
at 46.5 and 51.2 MeV/nucleon.
\end{abstract}


    The precise knowledge of the solar thermonuclear fusion of $^8$B
(from $^7$Be plus proton) is crucial for estimating the $^8$B solar 
neutrino flux and the predicted neutrino rates in terrestrial neutrino
measurements.
    The relevant $^7$Be(p,$\gamma$)$^8$B cross section $\sigma(E)$
is parameterized in terms of the astrophysical
factor $S_{17}(E)$ which is defined by 
$S_{17}(E) =\sigma(E) E \exp{[2\pi\eta(E)]}$ where
$\eta (E) = Z_1 Z_2 {\rm e}^2 /\hbar v$ is the Sommerfeld parameter. 
    The flux of $^8$B solar neutrinos is particularly important 
for the results of the 
Homestake, Super Kamiokande, and SNO experiments
\cite{Phs96}
which measure high-energy solar neutrinos mainly or solely 
from the $^8$B decay.

    Unfortunately, this cross section has not been known with sufficient 
accuracy for a long time, despite the fact that several comprehensive
direct measurements were reported for 
the $^7$Be(p,$\gamma$)$^8$B reaction
\cite{Parker68,Kavanagh69,Vaughn70,Filippone83,Hammache98}.
   The main difficulty in such experiments is
the determination of the effective target thickness
of the radioactive $^7$Be target.
    This difficulty is reflected in the fact that 
the results of these measurements 
can be grouped into two distinct data sets 
which agree in their energy dependence but disagree in their absolute
normalization by about 30\%.
    In view of this discrepancy,
experimental studies with different methods are highly desirable.

    As an alternative approach one can measure the inverse process, 
the  Coulomb dissociation (CD) of $^8$B into $^7$Be 
and proton~\cite{baurrebel96}.
     The CD yields are enhanced because thicker targets can be used 
and a larger phase space is available for CD.
    This method uses stable targets and thus is free from the difficulty 
of determining the effective target thickness.
    On the other hand, direct (p,$\gamma$) and Coulomb
dissociation measurements have different sensitivities to the multipole
composition of the photon fields.
    The E2 amplitude is enhanced in CD 
due to the large flux of E2 virtual photons, 
whereas it can be neglected in the (p,$\gamma$) reaction.

    Recently, Motobayashi {\it et al.} have performed a CD
experiment at $E(^8$B$)=46.5$~MeV/nucleon, yielding values 
for $S_{17}$ in the energy range 0.6$-$1.7~MeV \cite{Motobayashi94}.
    The extracted (p,$\gamma$) cross section is
consistent with the results from the lower group 
of direct-reaction data points \cite{Vaughn70,Filippone83,Hammache98}.
    Another measurement at 51.9 MeV/nucleon by the same group 
with improved accuracy
led essentially to the same conclusion~\cite{Kikuchi98}.

    In this article, we report on an experiment of the CD
of $^8$B at a higher energy of 254~MeV/nucleon performed at the SIS facility at 
GSI, Darmstadt, Germany.
    The present incident energy has several advantages compared to those
used in 
Refs. \cite{Motobayashi94,Kikuchi98}:
(i) due to the strong forward focusing of the reaction products 
the magnetic spectrometer KaoS~\cite{Senger93} was used for
a kinematically complete measurement with high detection efficiency
over a wider range of the p-$^7$Be relative energy;              
(ii)  effects that obscure the dominant contribution of E1 multipolarity
to the CD, such as E2 admixtures and higher-order
contributions, are reduced~\cite{Typel97,Bertulani94};
(iii) the M1 resonance peak at $E_{\rm rel} = 0.63$ MeV is excited 
stronger and therefore can be used to check the accuracy of the 
invariant-mass calculation.

    A $^8$B beam was produced by fragmentation of 
a 350~MeV/nucleon $^{12}$C beam from the SIS synchrotron impinging
on an 8.01 g/cm$^2$ beryllium target.
    The beam was isotopically separated in the fragment separator 
(FRS)~\cite{Geissel92} 
and transported to the standard target position of 
the spectrometer KaoS~\cite{Senger93}.
    The average beam energy of $^8$B at the center of the 
breakup target was 254.0~MeV/nucleon,
a typical $^8$B intensity was 10$^4$/spill with 7s extraction time.
    Beam particles were identified event by event 
with the TOF-$\Delta E$ method by using a plastic scintillator 
with a thickness of 5 mm placed 
68~m upstream from the target and a large-area scintillator wall
placed close to the focal plane of KaoS.
    About 20 \% of the beam particles were $^7$Be, 
which could however unambiguously be
discriminated from breakup $^7$Be particles by their time of flight.

    An enriched $^{208}$Pb target with a thickness of 199.7 ($\pm$ 0.2) 
mg/cm$^2$ and with an effective area of $20\times 22$ mm$^2$ 
was placed at the standard target position of KaoS.
    The reaction products, $^7$Be and proton, were analyzed 
by the spectrometer
which has a large momentum acceptance of $p_{\rm max}/p_{\rm min} 
\approx 2$ and
an angular acceptance of 140 and 280~mrad in horizontal and vertical 
directions, respectively.
    Two pairs of silicon micro-strip detectors, installed at about 
14 and 31~cm downstream from the target, 
measured the x- and y-positions and hence the scattering angles 
of the reaction products in front of the KaoS magnets.
    Each strip detector had a thickness of 300~$\mu$m, 
an active area of 56 $\times$ 56 mm$^2$, 
and a strip pitch of 0.1 mm.
    The energy deposited on each strip was recorded using an
analog-multiplexing technique.
    This configuration enabled us to measure 
opening angles of the reaction products with a 1$\sigma$-accuracy 
of 4.8 mrad which is mainly caused by angular straggling in the relatively 
thick Pb target.
    By reconstructing the vertex at the target with an 1$\sigma$-accuracy 
of 0.3 mm, 
background events produced in the target frame or in the strip detectors 
could be largely eliminated.  
    The remaining background events were found without the target 
under the same condition to be less than 0.5\% of the true events.

    Momenta of the reaction products were analyzed by 
trajectory reconstruction 
using position information from the micro-strip detectors and two  
two-dimensional multi-wire proportional chambers (MWPC) which 
detected the protons or the Be ions
close to the focal plane of KaoS with a position resolution
of about 1 mm.
   
    A large-area ($180 \times 40~{\rm cm}^2$) scintillator wall 
\cite{Senger93}
consisting of 30 plastic scintillator paddles with a
thickness of 2 cm each 
was placed just behind the MWPC.
    It served as a trigger detector
for the data acquisition system and as a stop detector for time-of-flight
(TOF) measurements.
    To distinguish breakup events from non-interacting beam particles,
the TOF wall was subdivided into two equal-size sections.
    The breakup events were characterized by 
coincident hits in the low-momentum (proton) and high-momentum ($^7$Be etc.) 
sections of the wall.  
    The beam normalization was done by analyzing $^8$B single hits on the
high-momentum section (downscaled by a factor of 1000) under the condition 
(from the tracking information) that
they originated from the active target area.
    From the measured angles and momenta of the breakup products 
the p-$^7$Be relative energy and the scattering angle $\theta_8$ of 
the center-of-mass of proton and $^7$Be (i.e. the excited $^8$B) with 
respect to the incoming beam were reconstructed.

\begin{figure}[tb]
\begin{center}
\vspace{-5mm}
\mbox{\epsfig{file=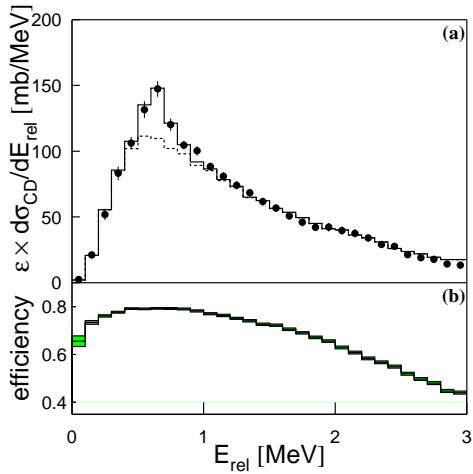,height=7cm}}
\end{center}
\caption[]{\footnotesize 
(a) Yields of breakup events (cross section $\times$ efficiency)
plotted as a function of relative energy.
The solid and dashed histograms denote simulated E1+M1 and E1 yields, 
respectively. 
(b) p-Be coincidence efficiency calculated by Monte Carlo simulations.}
\label{fi:relative}
\end{figure}

    The p-$^7$Be coincidence yield is shown in 
Fig.~\ref{fi:relative} (a) as a function of the p-$^7$Be relative energy.
    It is obvious from the figure that, contrary to the direct
measurements, the M1 resonance at $E_{\rm rel}=0.63$~MeV is not very 
pronounced. 
    This is related to the energy resolution of our experiment (see below),
and also to the relatively low sensitivity of CD to M1 
transitions.

    To evaluate the response of the detector system, 
Monte-Carlo simulations were performed using the code GEANT.
    The simulation took into account e.g. the finite size of the Si strip 
and MWPC detectors as well as our inability to discriminate proton and 
$^7$Be for very small opening angle where both particles hit the same 
strip.
    Events were generated with probabilities proportional to the
CD cross section calculated with a semi-classical formula
\cite{Baur86}.
    For the M1 resonance at $E_{\rm rel}=0.63$ MeV,
the (p,$\gamma$) cross section calculated from the total and gamma widths
measured by Filippone {\it et al.}~\cite{Filippone83} was used.
    The non-resonant contribution was obtained by normalizing 
the E1 (p,$\gamma$) cross section calculated by Bertulani~\cite{Bertulani96} 
with a scale factor of 1.20.  

    Further corrections in the simulation are due to
the feeding of the excited state at 429~keV in $^7$Be.
    We used the result by Kikuchi {\em et al.}~\cite{Kikuchi97}
who measured the $\gamma$-decay in coincidence with the
CD of $^8$B.

    The histograms in Fig.~\ref{fi:relative}(a) show 
the simulated E1+M1 yields (solid) and E1 yields (dashed).
   As seen in this figure, the shape and magnitude of the experimental
energy dependence are well reproduced.
This indicates that the CD yield is well described
by a combination of the M1 resonance and the pure E1 continuum.
    In the lower part of Fig.~\ref{fi:relative} we show that
the total efficiency calculated by the GEANT simulation is high
over the entire $E_{\rm rel}$ range covered in our study.
    From the Monte Carlo simulation we also  estimate our
relative-energy resolution to be e.g.
$\sigma(E_{\rm rel})=0.11$ and 0.22~MeV at $E_{\rm rel} = 0.6$ and 1.8~MeV,
respectively.

\begin{figure}[bt]
\begin{center}
\vspace{-5mm}
\mbox{\epsfig{file=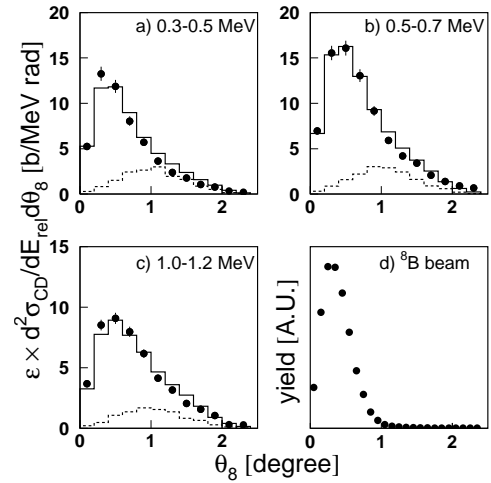,height=7cm}}
\end{center}
\caption[]{\footnotesize
(a,b,c)Yields of breakup events plotted against $\theta_8$, 
the scattering angle of the excited $^8$B,
for three relative energy bins.
The full histograms show the results of a simulation 
taking into account the measured angular spread of 
the incident $^8$B beam (shown in d) and assuming
E1+M1 multipolarity.
The dashed histograms show the simulated results for E2 contribution
according to the calculations of Bertulani\cite{Bertulani96}.
}
\label{fi:angular}
\end{figure}

   In order to estimate upper limits for a possible E2 contribution to
our yields, we have analyzed the $\theta_8$ distributions.
   Since we did not measure the incident angle of $^8$B at the target,
the experimental angular distributions represent the $\theta_8$ distributions 
folded with the angular spread of the incident
beam.
   In Fig.~\ref{fi:angular}, we plot the experimental yield
against the scattering angle $\theta_8$
for three relative-energy bins indicated in the figure;
the contribution from the measured beam spread 
is shown in Fig.~\ref{fi:angular} (d).  
   The full histograms represent
the results of a Monte-Carlo simulation for 
E1 excitation using the theory of Bertulani\cite{Bertulani96},
normalized to the experimental yield.
   The M1 transition also contributes to the 0.5$-$0.7 MeV bin
with the same angular dependence.
    The simulated E2 angular distributions are also shown by the dashed
histogram.
    The angular resolution was estimated to be 0.35 degrees  (1$\sigma$),
smaller than the observed widths.
    As seen in the figure
the experimental distributions are well reproduced by the simulation
using E1 and M1 multipolarities, and no room is left to add an additional
E2 component.
%
  The corresponding 3$\sigma$ upper limits for the E2/E1 transition amplitude
ratio are calculated from our model as $S_{\rm E2}$/$S_{\rm E1}=  
0.06\times 10^{-4}$, $0.3\times 10^{-4}$ and $0.6\times 10^{-4}$ 
for $E_{\rm rel}=0.3-0.5$, 0.5$-$0.7, and 1.0$-$1.2 MeV, 
respectively. 

    Recently, Esbensen and Bertsch have pointed out that interference between
E1 and E2 components could play an important role in CD
of $^8$B and should manifest itself in an asymmetry of the 
longitudinal-momentum distribution of $^7$Be\cite{Esbensen97}.
    Analyzing measured $^7$Be momentum spectra in term of this model, 
Davids {\it et al.} \cite{MSU2} have deduced 
$S_{\rm E2}$/$S_{\rm E1}=6.7^{+2.8}_{-1.9}\times 10^{-4}$ at 0.63 MeV
which is much higher than our upper limits above.
    One should keep in mind, however, that such a strong E2 component
need not necessarily affect the extraction of $S_{17}$: the dynamical
calculation of Esbensen and Bertsch \cite{Esbensen97} at 45.6 $A$ MeV 
shows that its
effect can be much smaller than what is obtained by simply adding E1 and E2
components.
Further experimental and theoretical studies are clearly desirable 
to clarify the magnitude of the E2 component in CD.

\begin{figure}[t]
\begin{center}
\vspace{-5mm}
\mbox{\epsfig{file=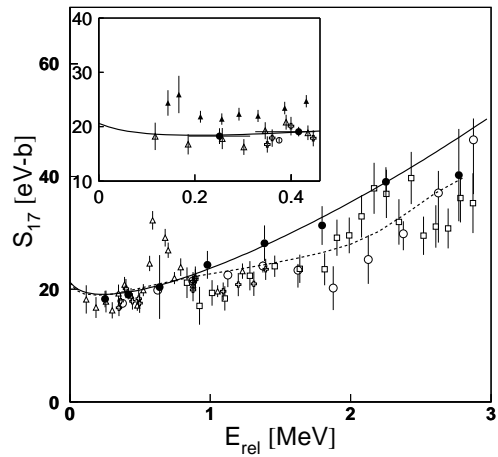,height=7cm}}
\end{center}
\caption[]{\footnotesize
 The astrophysical $S_{17}$ factors deduced from the present experiment
plotted as a function of the p-$^7$Be relative energy (closed circles),
in comparison with results from direct and other Coulomb-dissociation
experiments
(closed triangles from \cite{Kavanagh69}, 
open boxes from \cite{Vaughn70},
open triangles from \cite{Filippone83}, 
open crosses from \cite{Hammache98},
and open circles from \cite{Kikuchi98}).
  The solid curve shows the prediction of 
Bertulani~\cite{Bertulani96} fitted to our data,
while the dashed curve shows a fit of the theoretical curve
of Descouvemont {\it et al.}~\cite{DB94} to the combined 
datasets of Refs.~\cite{Vaughn70,Filippone83,Hammache98}.
  The insert shows the low-energy part of the figure, where we have fitted 
the energy dependence of Jennings {\it et al.}~\cite{Jennings98}
to our data. 
}
\label{fi:s17}
\end{figure}

   From the data and the simulation presented in Fig.~\ref{fi:relative} (a)
we can deduce the astrophysical $S_{17}$ factors for E1 multipolarity
by scaling the theoretical $S_{17}$ factors by the ratio of observed
and simulated counts.
    The resulting $S_{17}$ distribution is plotted in Fig.~\ref{fi:s17} as
a function of the p-$^7$Be relative energy, together with existing 
(p,$\gamma$)~\cite{Vaughn70,Filippone83,Hammache98} and 
CD \cite{Kikuchi98} results.
    The (p,$\gamma$) results were scaled to a peak cross section of the
$^7$Li(d,p)$^8$Li reaction of $147 \pm 11$~mb \cite{Adelberger98}. 
    The binning of our data was chosen to be approximately equal to the
FWHM of the $E_{\rm rel}$ resolution.
    The errors shown in Fig. 3 result from summing in quadrature statistical
errors and those resulting from uncertainties in the momentum calibration,
in the angular cutoff of pure Coulomb processes, and in the feeding of the
excited state in $^7$Be.
    Since we are mostly interested in the E1 component, 
the simulated M1 component of the $E_{\rm rel}=0.63$ MeV resonance was 
subtracted from the data.
    At low $E_{\rm rel}$,
our results agree well with the lower group of the (p,$\gamma$) results
\cite{Vaughn70,Filippone83,Hammache98} and the CD 
result of Kikuchi {\it et al.}~\cite{Kikuchi98}.
   We observe some discrepancies at higher $E_{\rm rel}$, however, as
shown in Fig.~\ref{fi:s17}.
   While we cannot explain the discrepancy with the RIKEN data 
\cite{Kikuchi97}, since
all of the assumptions underlying their and our analysis are identical,
the discrepancy to the direct measurements may be due to our neglect of
an E2 component which, in a proper theoretical treatment, might affect the
high-$E_{\rm rel}$ data points while leaving the low $E_{rel}$ data virtually
unchanged (see right-hand panel of Fig.11 in Ref.\cite{Esbensen97}).

    Several theoretical models for $^8$B have been proposed to predict
the shape and magnitude of the $S_{17}$ energy dependence, such as
one-body potential models 
(see e.g.~\cite{Typel97,Bertulani96} for recent calculations) or 
more complex many-body (cluster) models
(e.g.~\cite{DB94,Johnson92,Csoto95,Grigorenko}).
    Our data in Fig.~\ref{fi:s17} follow very closely the results 
obtained by Bertulani\cite{Bertulani96} 
over the entire range of energies as indicated by the solid curve
which was normalized to our data points using a scaling factor of 1.20. 
    The shape of the distribution predicted by the cluster model 
calculation~\cite{DB94}, which was favoured by the recent (p,$\gamma$) 
experiment of Hammache {\it et al.} \cite{Hammache98}, does not agree equally 
well with our results at the higher energies
(dashed line in Fig.~\ref{fi:s17}). 
    
    To obtain the zero-energy astrophysical $S_{17}$ factor, we follow
Jennings {\it et al.}~\cite{Jennings98} in fitting 
the theoretical energy dependence only to data points 
below $E_{\rm rel} \approx 0.45$ MeV (see insert in Fig. \ref{fi:s17})  
since this region should be largely free from uncertainties concerning 
nuclear excitations.
    From our two low $E_{\rm rel}$ data points we extracted 
$S_{17}(0) = 20.6 \pm 1.2 \pm 1.0$~eV-b where the first
contribution results from fitting the data points of Fig.~\ref{fi:s17}
and the second one is related to the uncertainty in extrapolating to 
zero $E_{\rm rel}$. This result
is compatible with the value of 
$S_{17}(0) = 19.0 \pm 1.0 \pm 0.2$ eV-b obtained by
Jennings {\it et al.} ~\cite{Jennings98} 
when fitting the combined data points of Refs. \cite{Filippone83,Hammache98},
 and also with the adopted value of $S_{17}(0)= 19^{+4}_{-2}$ eV-b from Ref.
~\cite{Adelberger98}.
  We note, however, that the question of possible backscattering 
losses of recoiling 
nuclei in direct (p,$\gamma$) and (d,p) reactions \cite{Weissman98,Strieder98}
is not yet solved.

  We conclude that we have demonstrated that high-energy Coulomb dissociation
is very useful for determining the astrophysical $S$-factor of the
$^7$Be(p,$\gamma$)$^8$B reaction at low energies.
  We have used an analytical formula from the literature~\cite{Jennings98}
to extrapolate our two low-energy $S$-factors to
$S_{17}(0)$ and obtain a value that is consistent with the most recent
compilation~\cite{Adelberger98}. 
  This supports essentially the standard-model prediction
for the high-energy solar neutrino flux \cite{Phs96}.

\bigskip
   
     We are grateful for the technical support by K.H.~Behr, A.~Br\"unle, and
K.~Burkard of the FRS staff. We thank N.~Kurz for help with the data 
acquisition and J.~Friese, R.~Gernh\"auser, E.~Badura, and J.~Hoffmann
for their support in designing and realizing the Si micro-strip
detector readout.
     Constant support and interesting discussions concerning the theoretical
aspects were contributed by S.~Typel, G.~Baur, B.J.~Jennings 
and H. Oberhummer.
     We thank P.~Descouvemont and D.~Baye for providing their 
numerical results.

\newpage


\begin{thebibliography}{10}

\bibitem{Phs96} 
J.N. Bahcall {\it et al.}, Phys. Lett. B {\bf 433}, 1 (1998).

\bibitem{Parker68} 
P.D.~Parker {\it et al.}, Astrophys. J. {\bf 153}, L85 (1968).

\bibitem{Kavanagh69} 
R.W.~Kavanagh {\it et al.}, Bull. Am. Phys. Soc. {\bf 14}, 1209 (1969).

\bibitem{Vaughn70} 
F.J.~Vaughn {\it et al.}, Phys. Rev. C {\bf 2}, 1657 (1970).


\bibitem{Filippone83} 
B. W.~Filippone {\it et al.}, Phys. Rev. C {\bf 28}, 2222 (1983).

\bibitem{Hammache98} 
F.~Hammache {\it et al.}, Phys. Rev. Lett. {\bf 80}, 928 (1998).

\bibitem{baurrebel96} 
G.~Baur and H.~Rebel, Annu. Rev. Nucl. Part. Sci. {\bf 46}, 321 (1996).

\bibitem{Motobayashi94}
T.~Motobayashi {\it et al.}, Phys. Rev. Lett. {\bf 70},  2680 (1994);
N.~Iwasa {\it et al.}, J. Phys. Soc. Japan {\bf 65}, 1256 (1996).

\bibitem{Kikuchi98}
T.~Kikuchi {\it et al.}, European Phys. J. A {\bf 3},  213 (1998). 

\bibitem{Senger93}
P. Senger {\it et al.}, Nucl. Instr. Meth. A {\bf 327}, 393 (1993).

\bibitem{Typel97}
S.~Typel and G.~Baur,  Phys. Rev. C {\bf 50}, 2104 (1994);
S.~Typel {\it et al.}, Nucl. Phys. A {\bf 613}, 147 (1997).

\bibitem{Bertulani94} 
C.A.~Bertulani, Phys. Rev. C {\bf 49}, 2688 (1994).

\bibitem{Geissel92}
H. Geissel {\it et al.}, Nucl. Instr. Meth. B {\bf 70}, 286 (1992).


\bibitem{Baur86}
C.A.~Bertulani and G.~Baur, Phys. Rep. {\bf 163} (1988) 300.

\bibitem{Bertulani96} 
C.A.~Bertulani, Z. Phys. A {\bf 356}, 293 (1996).

\bibitem{Kikuchi97}
T.~Kikuchi {\it et al.}, Phys. Lett. B {\bf 391}, 261 (1997), and
priv. comm. 

\bibitem{Esbensen97}
H.~Esbensen and G.F.~Bertsch, Nucl. Phys. A {\bf 600}, 37 (1997).


\bibitem{MSU2}
B.~Davids {\it et al.}, Phys. Rev. Lett. {\bf 81}, 2209 (1998).

\bibitem{Adelberger98} 
E.G.~Adelberger {\it et al.}, Rev. Mod. Phys. {\bf 70}, 1265 (1998).

\bibitem{DB94}
P.~Descouvemont and D.~Baye, Nucl. Phys. A {\bf 567}, 341 (1994).

\bibitem{Johnson92} 
C.W.~Johnson {\it et al.}, Astrophys. J. {\bf 392}, 320 (1992).

\bibitem{Csoto95}
A.~Cs\'ot\'o {\it et al.}, Phys. Rev. C {\bf 52}, 1130 (1995). 

\bibitem{Grigorenko}
L.V.~Grigorenko {\it et al.}, Phys. Rev. C {\bf 57}, R2099 (1998). 

\bibitem{Jennings98}  
B.K.~Jennings {\it et al.},
Phys. Rev. C {\bf 58} (1998) 3711.
 
\bibitem{Weissman98} 
L.~Weissman {\it et al.}, Nucl. Phys. A {\bf 630}, 678 (1998).

\bibitem{Strieder98} 
F.~Strieder {\it et al.}, Eur. Phys. J. A {\bf 3}, 1 (1998).

\end{thebibliography}
\end{document}